\documentstyle[aasms4]{article}
\input epsf
\begin{document}
\title{A Theoretical Calibration of $^{13}$CO LTE Column Density 
in Molecular Clouds.}

\author{Paolo Padoan}
\affil{Theoretical Astrophysics Center, 
       Juliane Maries Vej 30, DK-2100 Copenhagen, 
       Denmark}
\author{Mika Juvela}
\affil{Helsinki University Observatory, 
       T\"ahtitorninm\"aki, P.O.Box 14,
       SF-00014 University of Helsinki, Finland}
\author{John Bally}
\affil{Department of Astrophysics, Planetary, and Atmospheric Sciences,\\
       Center for Astrophysics and Space Astronomy,\\
       Campus Box 389, University of Colorado, Boulder CO 80309} 
\author{\AA ke Nordlund}
\affil{Astronomical Observatory and Theoretical Astrophysics Center, \\
       Juliane Maries Vej 30, DK-2100 Copenhagen, Denmark}

\authoremail{padoan@tac.dk}

\begin{abstract}

In this work we use models of molecular clouds (MC), and 
non-LTE radiative transfer calculations, to obtain a theoretical 
calibration of the relation between LTE $^{13}$CO column density
and true column density in MCs. The cloud models consist of 3 
dimensional grids of density and velocity fields obtained as 
solutions of the compressible magneto-hydrodynamic equations in 
a 128$^3$ periodic grid in both the supersonic and super-Alfv\'{e}nic 
regimes.  Due to the random nature of the velocity field and the 
presence of shocks,  the density spans a continuous range of values covering
over 5-6 orders of magnitude (from $\sim$0.1 to $\sim$10$^5$ cm$^{-3}$).  
As a result, the LTE column density can be calibrated over 3 orders 
of magnitude. We find that LTE column density of molecular clouds typically 
underestimates the mean $^{13}$CO true column density by a factor 
ranging from 1.3 to 7.  These results imply that the standard LTE
methods for the derivation of column densities from CO data 
systematically underestimate the true values independent of other
major sources of uncertainty such as the relative abundance of CO.

\end{abstract}


\section{Introduction}

Column densities in molecular clouds (MC) are estimated 
using the integrated antenna temperature of optically thin 
rotational transitions 
such as the J=1$\rightarrow$0 line of $^{13}$CO under the 
assumption of local thermodynamic equilibrium (Dickman 1978), 
adopting an empirical value for the [H$_2$]/[$^{13}$CO]
abundance ratio. The LTE calculations are based on a set of 
approximations that varies from work to work concerning the 
way to estimate excitation temperatures ($T_{ex}$), the partition 
functions ($Q$), and optical depths ($\tau$). The empirical 
determination of the [H$_2$]/[$^{13}$CO] abundance ratio 
relies on the LTE calculations, and on the empirical relation 
between gas column density and stellar extinction (Lilley 1955; 
Jenkins \& Savage 1974; Bohlin, Savage \& Drake 1978).
The conversion of the LTE $^{13}$CO column density into 
total gas column density suffers from several uncertainties 
that we do not discuss here. 

In this paper, we show that the true $^{13}$CO column density 
is underestimated by the usual LTE approximations.  
Although previous works have indicated that the LTE 
approximations are good (eg Dickman 1978; Park, 
Hong \& Minh 1996), real MCs have structure and kinematics 
that are far more complex than assumed in those works.  The 
effect of complexity on the resulting spectra is investigated 
here.  Juvela (1997), using more realistic density fields
created with fractal models or structure tree statistics, 
has shown that the estimation of column densities 
is uncertain if the density structure of the cloud is unknown.

Padoan et al. (1997a) produced a catalog of 
artificial MCs containing grids of 90$\times$90 spectra 
of different molecular transitions for a total of more than 
one million spectra. The spectra were obtained with a non-LTE 
Monte Carlo code (Juvela 1997) starting from density and velocity 
fields that provide realistic descriptions of the observed 
physical conditions in MCs.  The density and velocity fields 
are obtained as solutions of the magneto-hydrodynamic (MHD) 
equations in a 128$^3$ periodic grid and in both 
super-Alfv\'{e}nic and highly supersonic regimes of random 
flows (Padoan \& Nordlund 1997). The resulting 
density fields span a continuous range of values from 0.1 to 
10$^5$ cm$^{-3}$ which produce column densities ranging
over three orders of magnitude or more. These cloud models have been 
shown to reproduce the observed statistical properties of MCs 
(Padoan, Jones \& Nordlund 1997, Padoan \& Nordlund 1997)
and are realistic enough to allow a theoretical calibration of 
the $^{13}$CO column density.    

For the purpose of this work we use six cloud models from the 
catalog of artificial clouds by Padoan et al.\ (1997a). We use 
90$\times$90 grids of spectra of J=1$\rightarrow$0 $^{13}$CO 
and $^{12}$CO lines from three 5~pc diameter artificial clouds 
and from three 20~pc clouds. For a detailed description of the 
construction of the synthetic molecular maps we refer the reader 
to Padoan et al. (1997a).  The structures seen in these simulations
result entirely from the random character of the flows.  The
influence of gravity and heating effects of external radiation
fields are ignored in these MHD calculations.

\section{Excitation Temperatures}

Over most of the calculated volume, the $^{13}$CO rotational 
transitions are sub-thermally excited in our artificial clouds. 
The excitation temperature, $T_{ex}$, is higher in regions of 
larger gas density.  Fig.~1 shows three slices of the 
3-D datacubes illustrating typical images of $T_{ex}$ 
of the J=1$\rightarrow$0 $^{13}$CO and gas density in the 5~pc 
cloud model. As expected, $T_{ex}$ correlates with gas density.  
The distribution of $T_{ex}$ has a smaller dynamic range than the 
density distribution.  The intensity scale in the 
density map is proportional to the logarithm of the density in 
order to compress the contrast.  Fig.~2 shows scatter plots of 
$T_{ex}$ versus logarithmic gas density, $Log(n)$.  The horizontal 
dashed line represents the mean excitation temperature $<T_{ex}>$.
$<T_{ex}>\approx 4 K$ in both cloud models.  However, the 20~pc 
models are a bit `colder' because they have larger regions of very 
low density gas than the 5~pc models (see Padoan et al. 1997a). 
The J=1$\rightarrow$0 transition of $^{13}$CO is therefore
strongly sub-thermally excited. 

In the following section we estimate the LTE column density for these
models.  We estimate the excitation temperature from the 
peak temperature of the J=1$\rightarrow$0 transition of $^{12}$CO, 
at every spectrum in the 2--D grids, adopting the abundance ratio 
[$^{12}$CO]/[$^{13}$CO]=50. The mean value of that temperature 
is plotted in Fig.~2 as a continuous horizontal line. It is apparent 
that the estimated value of $T_{ex}$ is about twice as large as 
the correct one.

\section{The LTE and True Column Density}

The J=1$\rightarrow$0 transitions of $^{13}$CO and $^{12}$CO can 
be used to estimate the column density of $^{13}$CO, under the 
assumption of local thermodynamic equilibrium (LTE). Different 
authors made use of different approximations, when performing 
the LTE calculations to estimate column densities from observational 
data. Therefore we will determine LTE column densities from our
models in several different ways to reproduce the various methods 
presented in the literature.  We briefly summarize the formulae that are 
used in the LTE calculations (eg Dickman 1978; Harjunp\"{a}\"{a} \& 
Mattila 1996) and we then add some simplyfing assumptions. 

The two basic assumptions are: (1) The excitation temperature is 
uniform along the line of sight. (2) The J=1$\rightarrow$0 
excitation temperature of $^{13}$CO and $^{12}$CO are the same. 
The excitation temperature, $T_{ex}$, is estimated 
from the formula:

\begin{equation}
T_R=[J(T_{ex})-J(T_{bg})](1-e^{-\tau})
\label{1}
\end{equation}
where $T_{bg}=2.7K$ is the background temperature, $\tau$ is the optical 
depth, $T_R$ is the radiation temperature, and the function $J(T)$ is:

\begin{equation}
J(T)=\frac{T_0}{exp(T_0/T)-1}
\label{2}
\end{equation}
where $T_0=h\nu_{10}/k$ and $\nu_{10}$ is the frequency of the 
transition J=1$\rightarrow$0 of $^{13}$CO. $T_{ex}$ is estimated 
from the formula (\ref{1}), where the peak radiation temperature 
of the J=1$\rightarrow$0 transitions of $^{12}$CO
is used for $T_R$, and the same transition is assumed to 
be optically thick
($\tau \gg 1$). Then equation (\ref{1}) can be used again, 
with the estimated value of $T_{ex}$, to determine $\tau$ 
in each channel, $\tau_{ch}$.  The column density of $^{13}$CO 
in the ground state, $N_0$, is given by:

\begin{equation}
N_0=6.39\times10^{14}\frac{\sum \tau_{ch}\Delta v}{1-e^{-T_0/T_{ex}}}cm^{-2}
\label{3}
\end{equation}
where $\Delta v$ is the channel width in km~s$^{-1}$. To obtain 
the total column density of $^{13}$CO, $N_0$ must be multiplied by 
the partition function $Q$:

\begin{equation}
N_1=N_0\times Q=N_0\times\sum_{J=0}^{J_{max}}(2J+1)exp\left[\frac{-h\nu_{10}J(J+1)}{2kT_{ex}}\right]
\label{4}
\end{equation}

To use equation (\ref{4}), it is usually assumed that 
the same $T_{ex}$ is valid for all rotational 
states of the molecule. The formulae are taken from 
Harjunp\"{a}\"{a} \& Mattila (1996), and we call $N_1$ the 
LTE column density estimated in this way. The optically thin limit 
($\tau\ll 1$) is sometimes adopted (eg Park, Hong \& Minh 1996), 
which corresponds to the approximation: $(1-e^{-\tau})\approx \tau$. 
We call $N_2$ the LTE column density estimated under
this assumption. Another frequently adopted  simplification 
(eg. Dickman 1978; Park, Hong \& Minh 1996) is the following 
approximation to the partition function:

\begin{equation}
Q\approx 2T_{ex}/T_0 
\label{5}
\end{equation}
(Penzias, Jefferts \& Wilson 1971). We call $N_3$ the LTE column density 
estimated like $N_1$, but with the approximation (\ref{5}). In 
some work (eg Dickman 1978), equation (\ref{3}) is not evaluated as 
the sum of $\tau_{ch}\Delta v$ over all channels, but rather using the 
line center optical depth multiplied by the line full width at 
half maximum (FWHM). We call this case $N_4$, where the partition 
function is also approximated with the formula (\ref{5}). Finally, 
$N_5$ is calculated as $N_4$, but the FWHM and the peak temperature 
are that of a Gaussian fit to the line profile. 

All results are summarized in Table~1, where the LTE column densities 
$N_i$ are compared with the true column density $N$, that is the 
$^{13}$CO column density of the model clouds. The value of $<N_i/N>$ 
illustrates the typical error made with the LTE approximation. 
On the average, the most detailed LTE calculations yield a value 
of $N_1$ that is about 65\% of the true column density in the 5~pc cloud 
model, and 40\% of $N$ in the larger scale 20~pc cloud. LTE column densities 
estimated with additional approximations underestimate the true 
column densities even more with  the worst case being the optically 
thin approximation, $N_2$, although the J=1$\rightarrow$0 transitions of 
$^{13}$CO is optically thin in most of the map.  The value 
of $<N_i>/<N>$ is an estimator of the ratio of total estimated mass and real 
total mass of the cloud. The best case, $N_1$, yields 50\% and 72\% 
of the total mass of 20~pc and 5~pc models respectively; the worst case, 
$N_2$, yields 37\% and 56\% of the total mass for the same models. 
The estimated mean excitation temperature, $<T_{ex}>$, is lower than 
the kinetic temperature ($T_K=10K$), but higher than the true mean 
excitation temperature of the model clouds (cf Fig.~2).  In Fig.~3 
the probability distributions of $N_1/N$ are plotted for both scales.

In Fig.~4, the value of $<N_1>$ is compared with the LTE column density 
estimated in the same way as $N_1$, but assuming a given constant 
$T_{ex}$ of 4, 6, 8, and 10 K. If very low $T_{ex}$ are adopted, 
the LTE column density can grow to the point of overestimating 
the true column density (cf Harjunp\"{a}\"{a} \& Mattila 1996, Fig.~2). 
If instead, the value adopted for the constant $T_{ex}$ is close to the 
value of $<T_{ex}>$ estimated with the peak temperature of $^{12}$CO, 
the constant temperature column densities underestimate $N$ more than 
$N_1$ does. Moreover, the scatter around the estimated mean column 
density is always larger than for $N_1$.

\section{Theoretical Calibration of $^{13}$CO LTE Column Density}

Fig.~5 and Fig.~6 are log-log plots of $N_1$ versus the true 
column density $N$.  Our models span three orders of magnitude in 
column density, a much larger range than can be sampled by observations 
when estimating gas column density using stellar extinction determinations 
(eg. Encrenaz, Falgarone \& Lucas 1975; Tucker et al.\ 1976; Dickman 1978;
Dickman \& Herbst 1990; Lee, Snell \& Dickman 1991; Lada et al.\ 1994; 
Harjunp\"{a}\"{a} \& Mattila 1996). As discussed in the introduction,
our cloud models have been shown to be consistent with the known properties
of MCs.  Therefore, we use these models to determine the scale factor $f$
that can be used to convert LTE based column density estimates to a more precise
estimate of the true column density. Unfortunately it is rather difficult 
to fit the relations $N-N_i$ with simple mathematical functions.  Instead, 
we plot the calibration factors, $f(N_i)=N/N_i$, for $N_1$, $N_2$, 
and $N_4$, in Fig.~7. When the LTE column density of $^{13}$CO is estimated 
from observational spectra, it should be multiplied by the calibration 
factor $f(N_i)$. The LTE column densities perform better at larger 
density, apart from $N_2$ that is good only in the optical thin limit. 
The often used and simplest LTE estimate, $N_4$, is rather good at large 
density, but it is up to 7 times smaller than the true column density, at low 
densities.

\section{Discussion}

Several authors have produced synthetic molecular spectra using simple models 
for the structure and kinematics of MCs (eg Zuckerman \& Evans 1974; 
Leung \& Liszt 1976; Baker 1976; Dickman 1978; Martin, Hills \& Sanders 1984; 
Kwan \& Sanders 1986; Albrecht \& Kegel 1987; Tauber \& Goldsmith 1990; 
Tauber, Goldsmith \& Dickman 1991; Wolfire, Hollenbach \& Tielens 1993; 
Robert \& Pagani 1993; Park \& Hong 1995; Park, Hong \& Minh 1996; Juvela 1997). 
Leung \& Liszt (1976) used a single micro-turbulent cloud and Dickman (1978) 
a collapsing 
spherical cloud.  Models of many clumps moving at random velocities larger than 
their intrinsic line-width (macro-turbulence) were used by Zuckerman \& Evans (1974)
and by Baker (1976). Macro-turbulence was recognized to be necessary to obtain 
centrally peaked CO line profiles. Macro-turbulent clumpy models include the 
use of different velocity and clustering laws (eg Kwan \& Sanders 1986),
increasing clump filling factor towards the center of the cloud 
(eg Tauber et al. 1991), variations of correlation length in the clump 
distribution (Albrecht \& Kegel 1987), a description of chemical reactions and 
heating (eg Wolfire et al. 1993), a low density warmer inter-clump medium 
(Park et al. 1996), density fields generated with fractal models or structure 
tree statistics (Juvela 1997).
 
Although clumpy models have being useful to study the properties of 
MCs, they can only provide a schematic representation of the structure and 
kinematics of MCs, since their velocity and density fields are not solutions
of the fluid equations. MCs are highly dynamical objects with a 
continuous range of gas density values and therefore a 
fluid description is appropriate and necessary. Stenholm \& Pudritz (1993) 
and Falgarone et al. (1994) produced molecular spectra from fluid models 
of clouds.  Stenholm \& Pudritz (1993) did not solve the MHD equations, but rather 
used a sticky particles code with an imposed spectrum of Alfv\'{e}n waves
(Carlberg \& Pudritz 1990). They computed molecular spectra under the simple 
assumption of LTE. Falgarone et al.\ (1994) did not solve the radiative 
transfer equations, but simply calculated density weighted radial velocity 
profiles. They used a very high resolution turbulence simulation 
(Porter, Pouquet \& Woodward 1994), but with a low Mach number ($\le 1.1$)
and without magnetic fields. 
Though these cloud models are solutions of the fluid equations, 
they are not a realistic representation of the physical 
conditions in MCs, because of the low Mach numbers and of the low
density contrasts. 

In the present work  we make use of highly supersonic MHD turbulence simulations
where the density fields span a continuous range of values covering about 
six orders of magnitude.  We perform the radiative transfer calculations with 
a non-LTE Monte Carlo code, producing high resolution molecular maps 
(90$\times$90 spectra) for different molecular transitions. 
We have already shown in other papers that our cloud models are excellent 
description of the observed physical conditions in MCs (Padoan, Jones \& Nordlund
1997; Padoan \& Nordlund 1997; Padoan et al.\ 1997a; Padoan et al.\ 1997b).

Our main result is that the {\it LTE $^{13}$CO column density of MCs underestimates
the true column density by a factor of 1.3 to 7.} Although previous works 
have shown that the LTE approximations are good (eg. Dickman 1978;  
Park, Hong \& Minh 1996),  these authors found that in regions of 
low density, the LTE column density is underestimated. Our results are consistent 
with previous studies.  In our more realistic cloud models the density has 
a continuous distribution and a large fraction of the volume is at 
very low densities where conditions are far from LTE. As an example,
in our models, only 20-30\% of the total mass resides in regions 
where the density is larger than in the clumps modeled by Park et al. (1996).   
On the other hand, the peak density in our
model is larger than in Park et al. (1996) model. 

In Fig.~8 we show that three basic assumptions of the LTE calculations 
are not correct. The probability distribution functions of $T_{ex}$ of 
different transitions show that: (1) $T_{ex}$ values are considerably 
smaller than the kinetic temperature (10~K); (2) $T_{ex}$ of $^{13}$CO 
and $^{12}$CO are quite different; (3) the $T_{ex}$ along a single line of 
sight is not uniform, but has a broad distribution of values. It is 
not surprising therefore to find discrepancies between LTE column
densities and true column densities.

In this work we have treated the small scale (5~pc) and large scale 
(20~pc) simulations separately. The distinction is important, because 
the volume filling factor of dense regions grows with decreasing scale. 
The peak density in a simulation is independent of simulation scale but
the mean density is lower in the larger simulations where the density contrast 
is larger.  Larger scales have stronger turbulent motions that cause 
a larger density contrast compared with smaller scales. 
Therefore, the calibration of LTE column densities, 
using a realistic cloud model, requires consideration of the length 
scales involved. 

Observations have demonstrated that external radiation fields tend
to make cloud exteriors warmer than their interiors (cf. Castets et al 1990).
The $^{12}$CO cloud `photosphere' (where the line optical depth reaches unity)
tends to lie in a warmer region than the portion of the cloud where the
$^{13}$CO emission is produced.  Thus, estimates of T$_{ex}$ based on 
$^{12}$CO tend to overestimate its true value.  This leads to an 
additional underestimation of the column density.  

Both the MHD and radiative transfer calculations adopted a constant
temperature. Heating and cooling processes are assumed to exactly
balance at one specific temperature.
If the thermal balance and temperature were computed
self-consistently, the results of the present calculations could
be slightly altered.

\acknowledgements

This work has been partially supported by the Danish National Research Foundation
through its establishment of the Theoretical Astrophysics Center.
Computing resources were provided by the Danish National Science Research Council,
and by the French `Centre National de Calcul Parall\`{e}le en Science de la Terre'.
PP is grateful to the Center for Astrophysics and Space Astronomy (CASA), in
Boulder (Colorado), for the warm hospitality offered during the period in which
this paper has been written.
The work of MJ was supported by the Academy of Finland Grant No. 1011055.

\newpage
\begin{figure}
\centering
\leavevmode
\epsfxsize=1.0
\columnwidth
\epsfbox{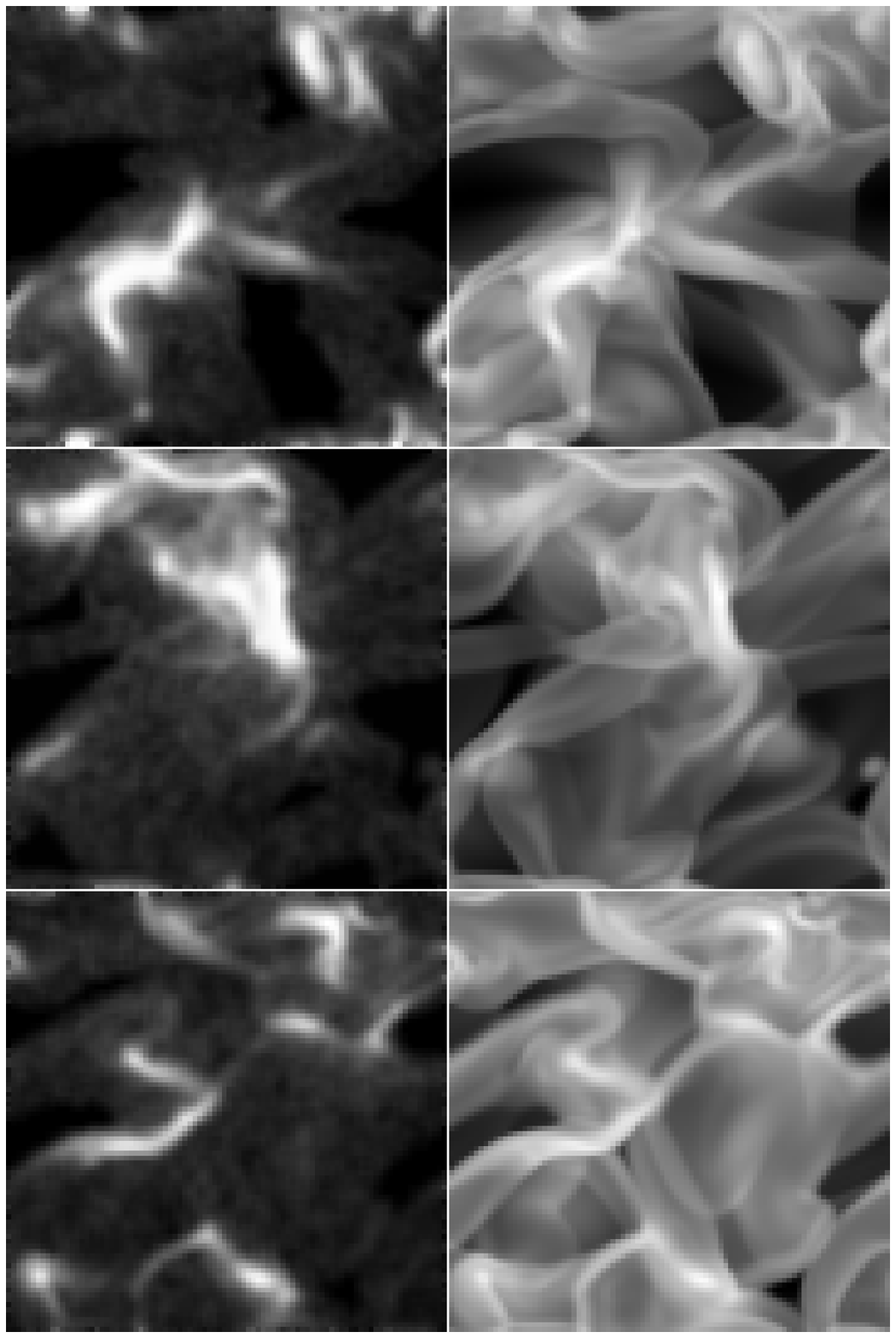}
\vspace{5cm}
\caption[]{2-D slices of excitation temperature of $^{13}$CO, J=1$\rightarrow$0
(left hand side panels), and of the logarithm of the gas density (right hand
side panels). The excitation temperature is larger in regions of larger density.}
\end{figure}

\newpage
\begin{figure}
\centering
\leavevmode
\epsfxsize=0.5
\columnwidth
\epsfbox{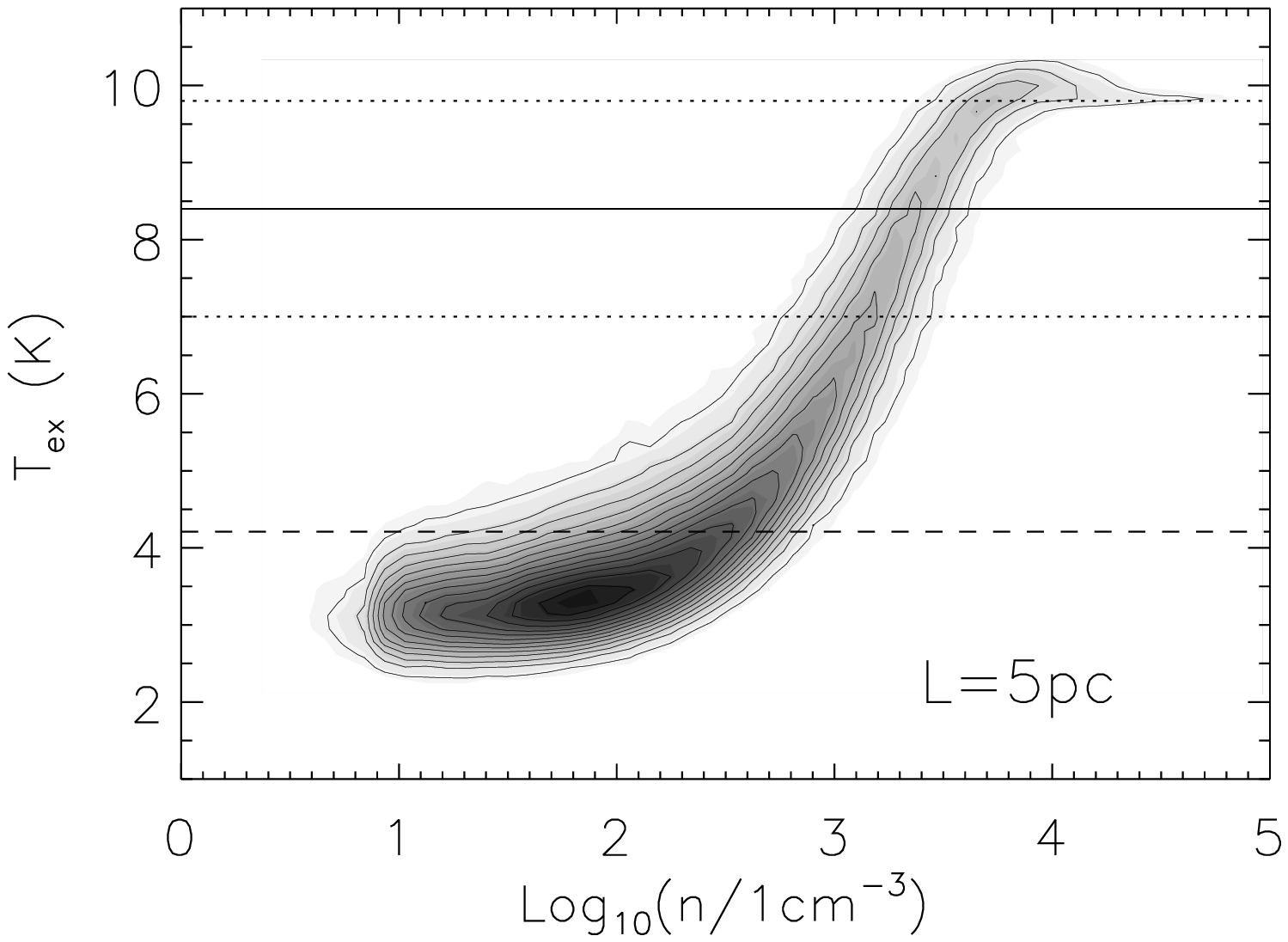}
\epsfxsize=0.5
\columnwidth
\epsfbox{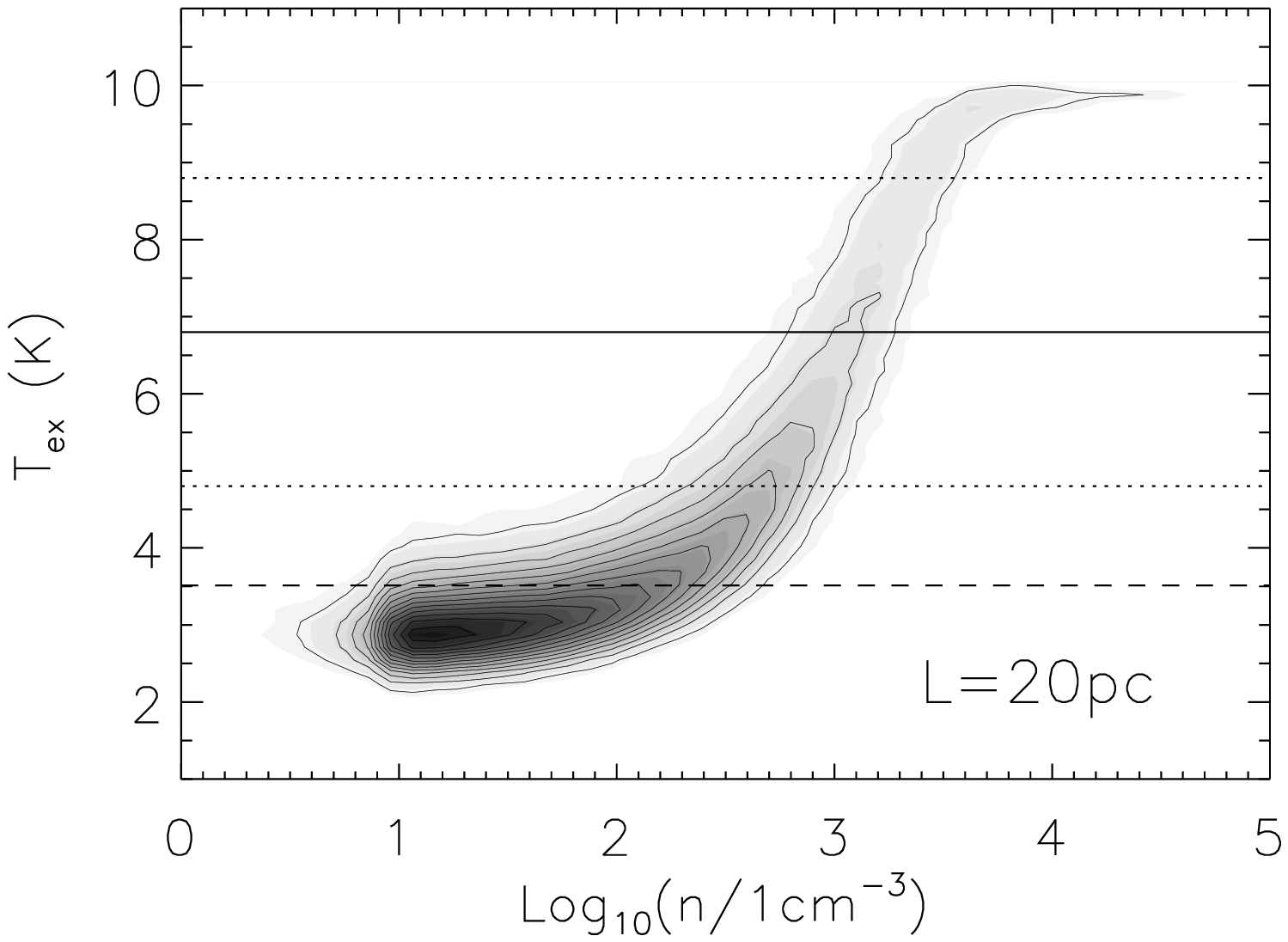}
\caption[]{Scatter plot of excitation temperature of $^{13}$CO, 
J=1$\rightarrow$0, versus the logarithm of the gas density, in 5pc 
(left) and 20pc (right) cloud models. The dashed horizontal line 
marks the mean excitation temperature. The continuous horizontal line
marks instead the mean excitation temperature estimated in the LTE
calculations to determine the LTE column density N$_1$ (the dotted lines
illustrate the $1-\sigma$ values).}
\end{figure}

\newpage
\begin{figure}
\centering
\leavevmode
\epsfxsize=1.0
\columnwidth
\epsfbox{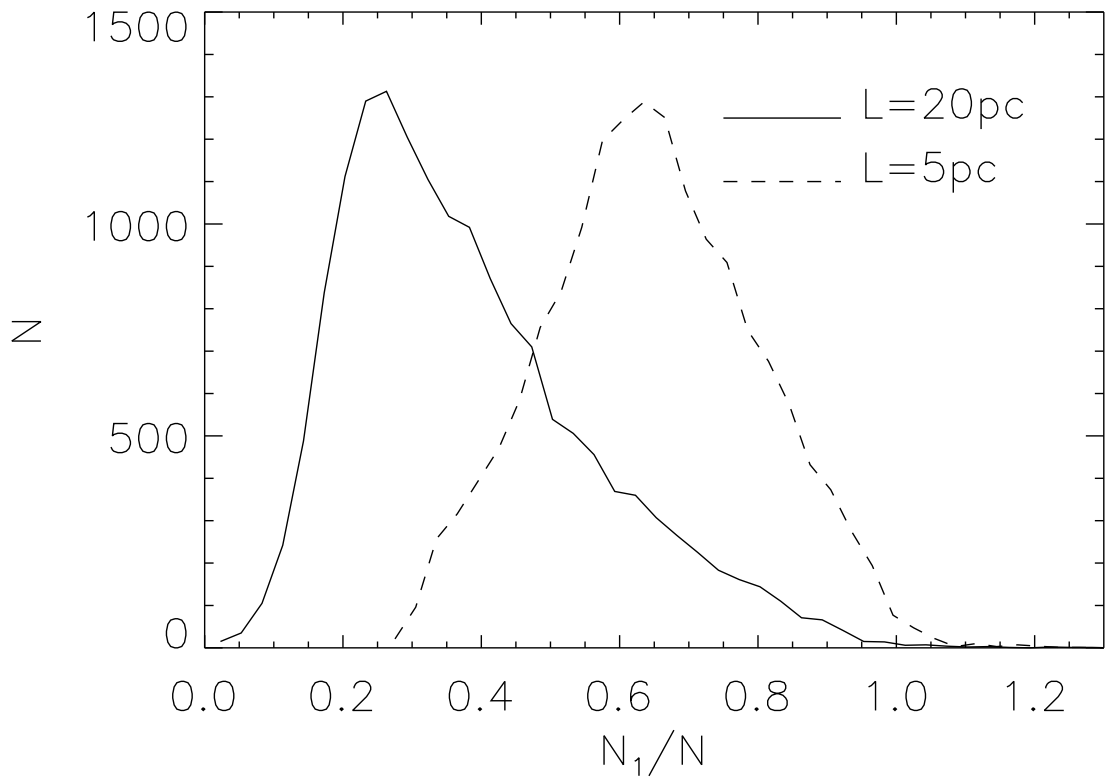}
\caption[]{Probability distributions of the best LTE column density estimate
divided by the true column density, $N_1/N$.}
\end{figure}

\newpage
\begin{figure}
\centering
\leavevmode
\epsfxsize=0.5
\columnwidth
\epsfbox{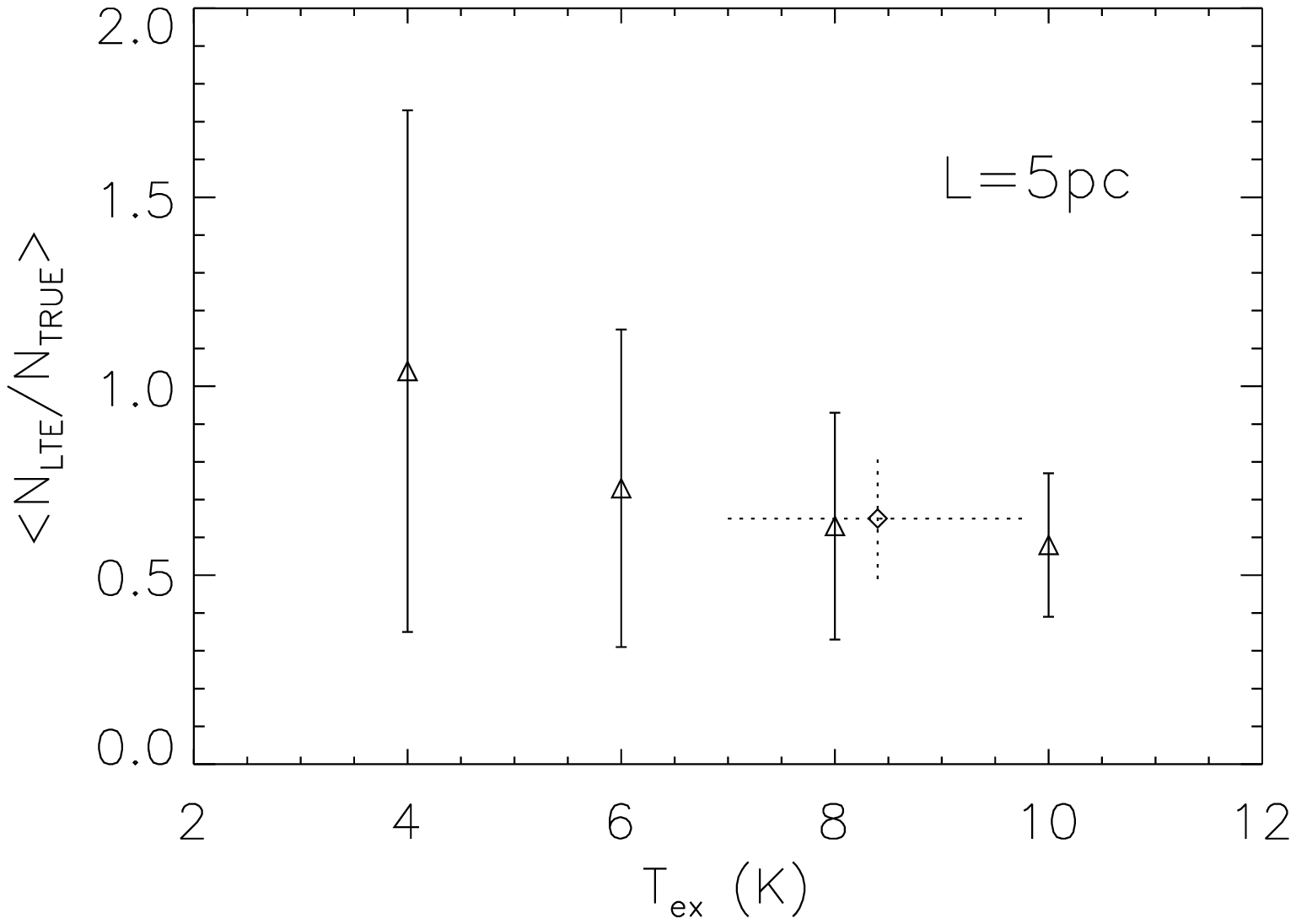}
\epsfxsize=0.5
\columnwidth
\epsfbox{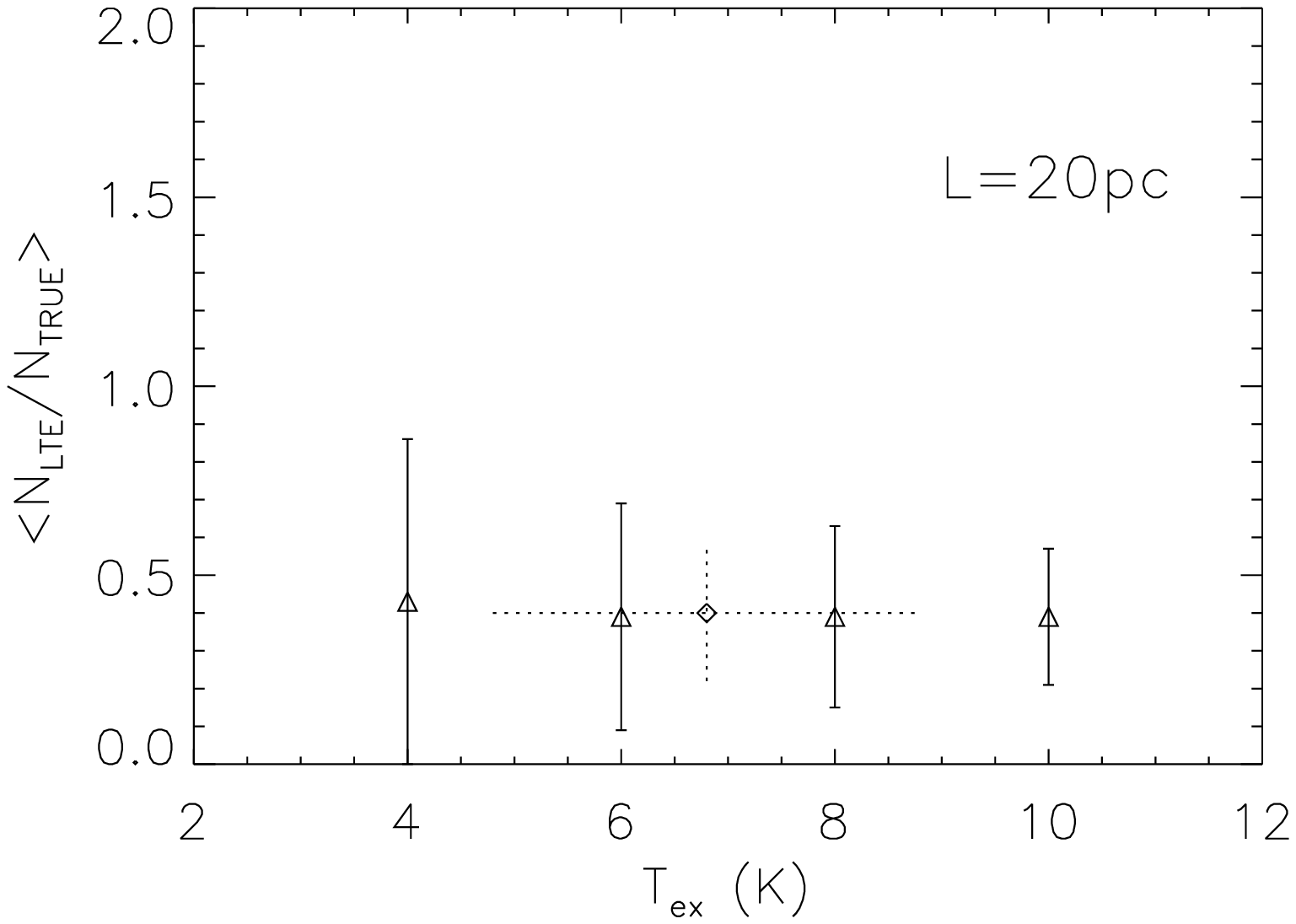}
\caption[]{Estimated LTE column densities, using the same assumptions as in N$_1$,
but with a constant T$_{ex}$. The bars represent the $1-\sigma$ dispersion around
$<N_{LTE}/N_{TRUE}>$. The diamond symbol marks the value of $<N_1/N>$ and the
mean  excitation temperature, $<T_{ex}>$, estimated to determine N$_1$.}
\end{figure}

\newpage
\begin{figure}
\centering
\leavevmode
\epsfxsize=1.0
\columnwidth
\epsfbox{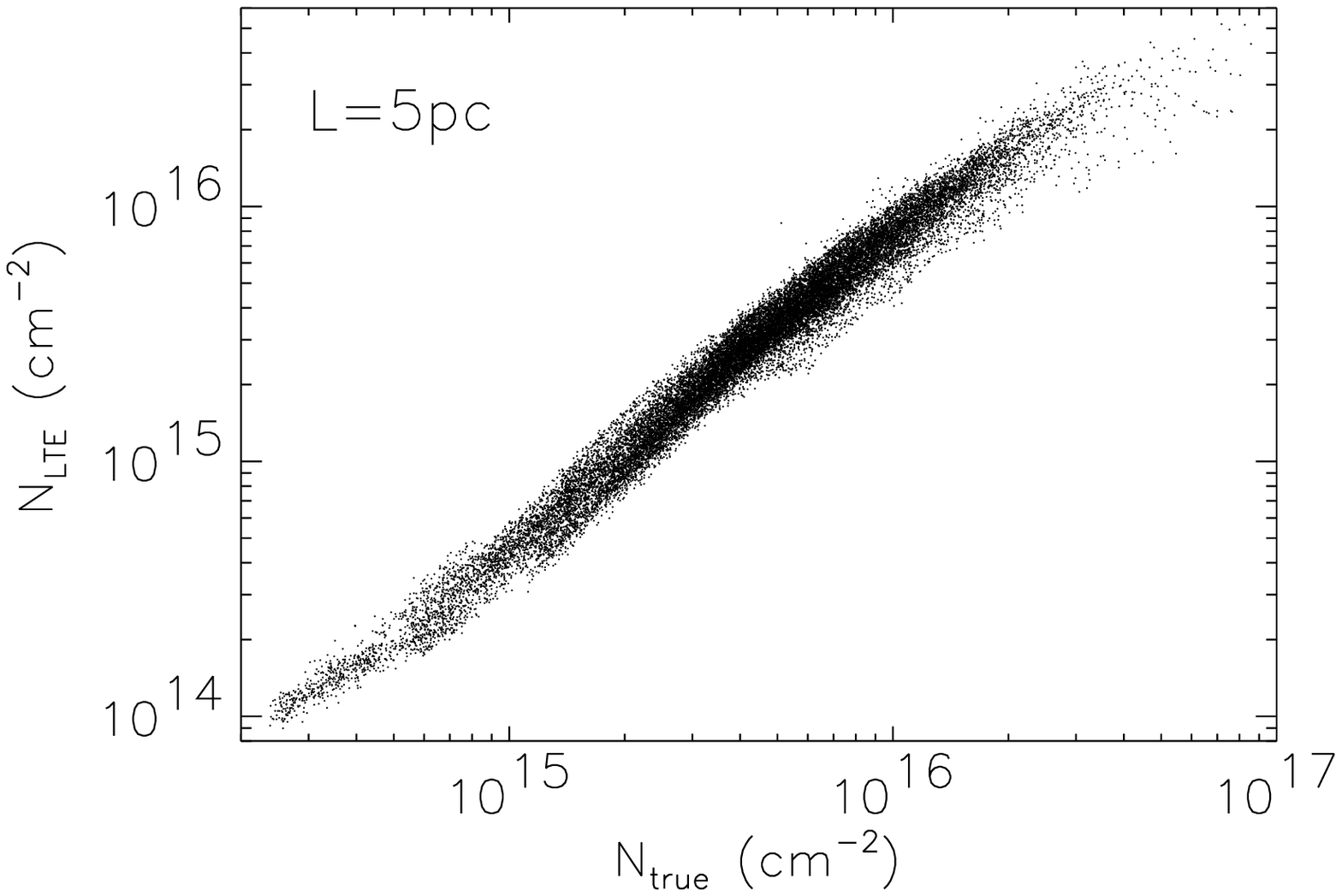}
\caption[]{LTE column density of $^{13}$CO, versus the true column density, for
5pc cloud models. LTE always underestimate the true column density.}
\end{figure}

\newpage
\begin{figure}
\centering
\leavevmode
\epsfxsize=1.0
\columnwidth
\epsfbox{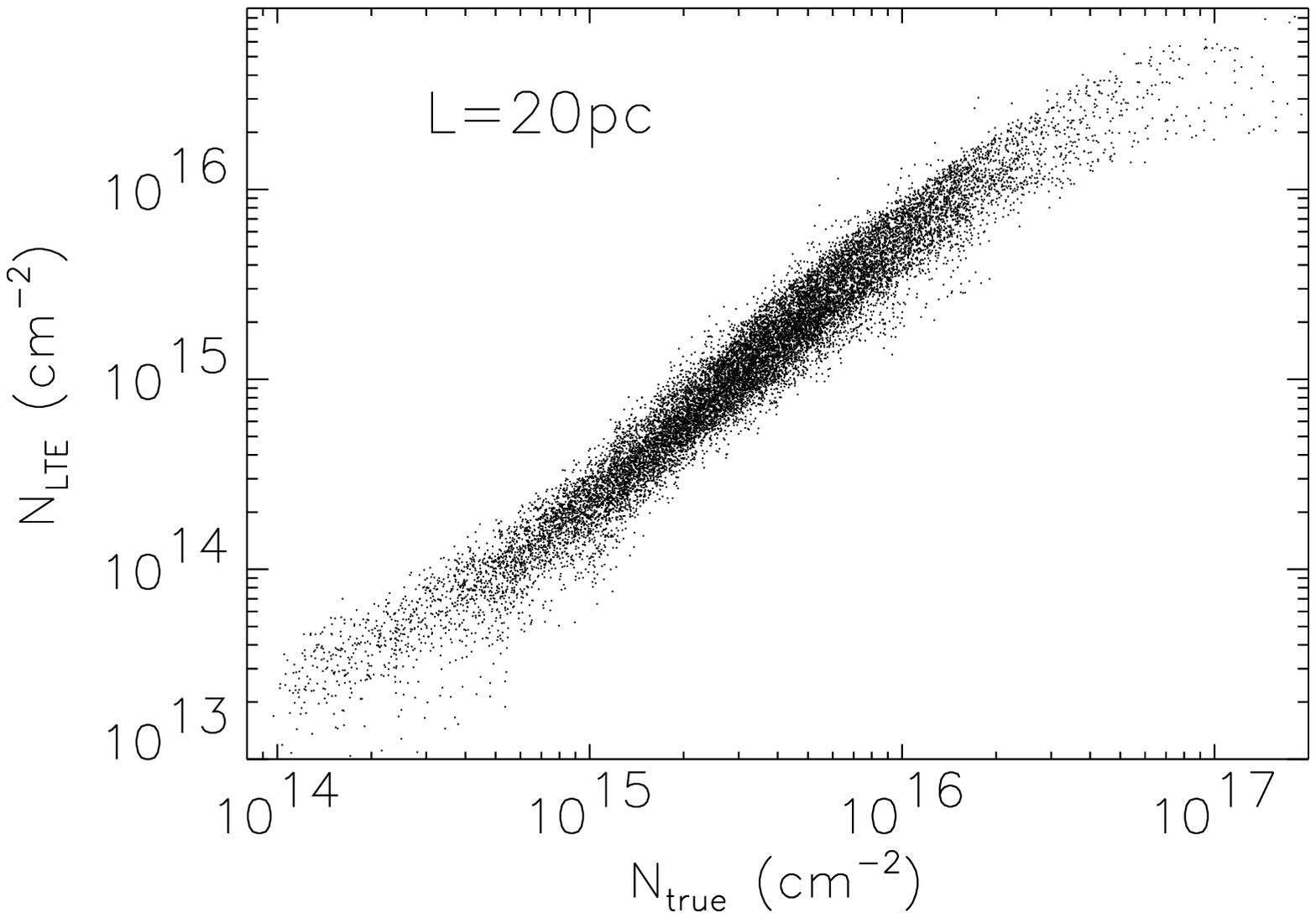}
\caption[]{LTE column density of $^{13}$CO, versus the true column density, for
20pc cloud models. LTE always underestimate the true column density.}
\end{figure}

\newpage
\begin{figure}
\centering
\leavevmode
\epsfxsize=0.5
\columnwidth
\epsfbox{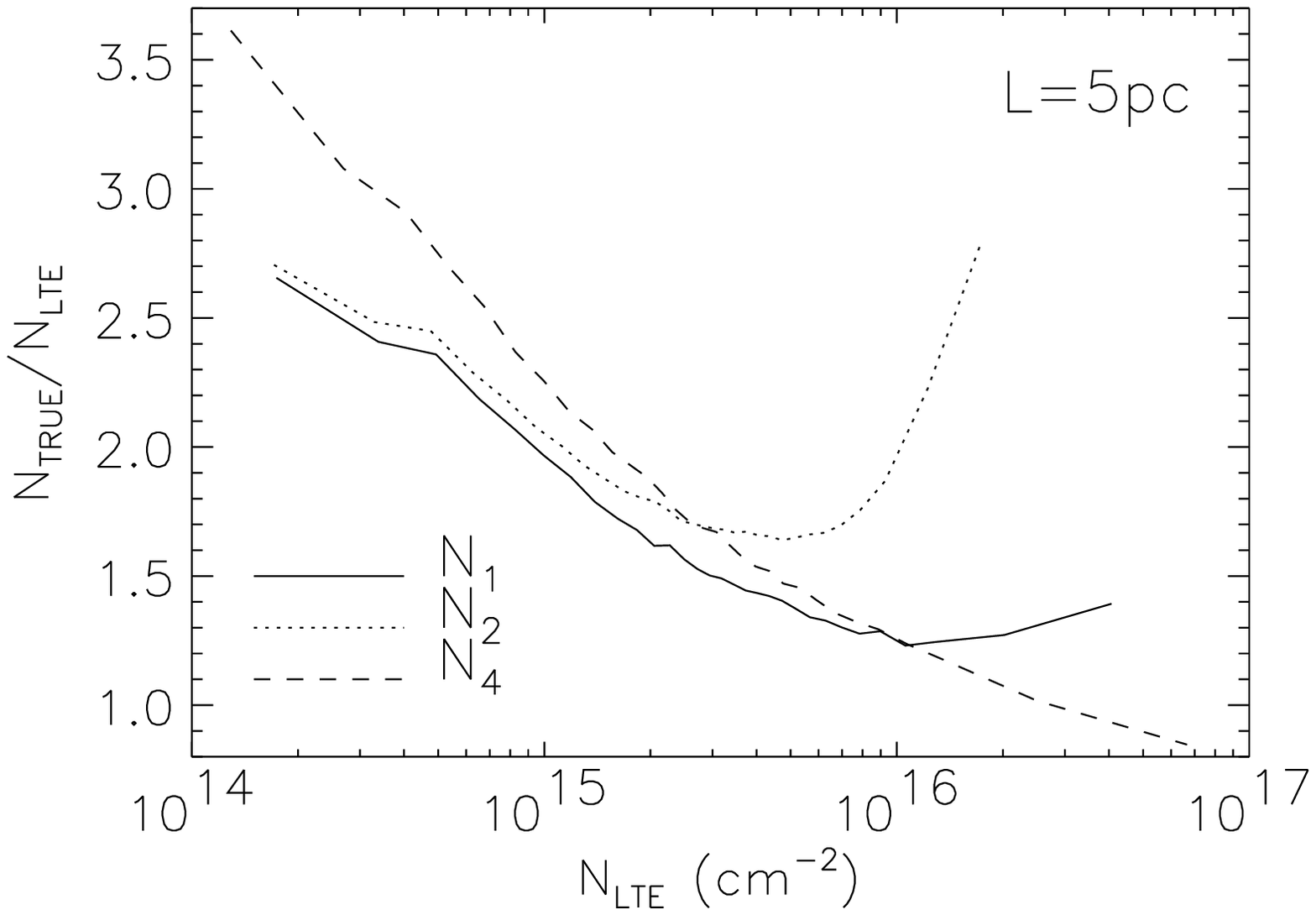}
\epsfxsize=0.5
\columnwidth
\epsfbox{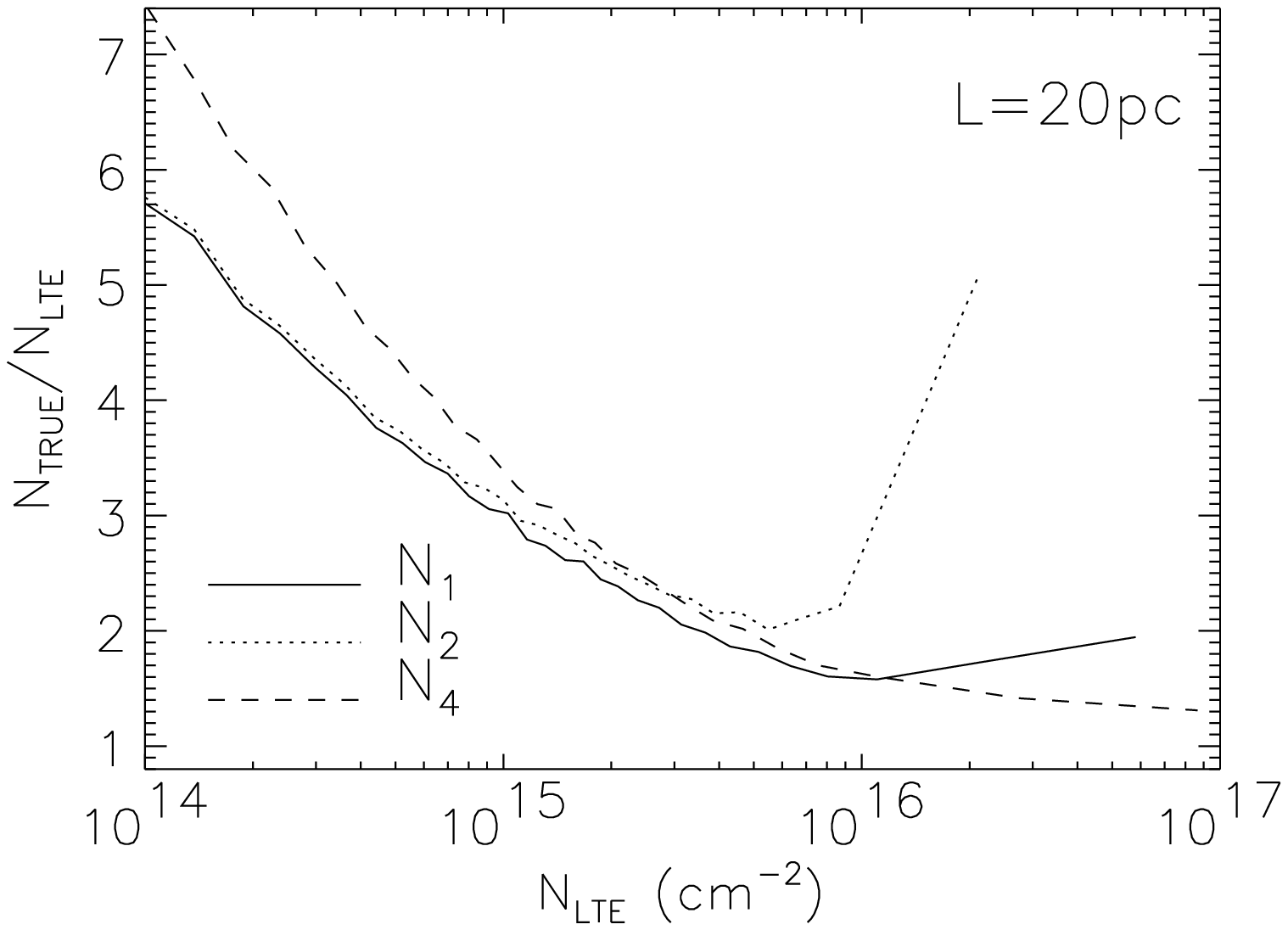}
\caption[]{Ratios of true and LTE column density, versus LTE column density.}
\end{figure}

\newpage
\begin{figure}
\centering
\leavevmode
\epsfxsize=1.0
\columnwidth
\epsfbox{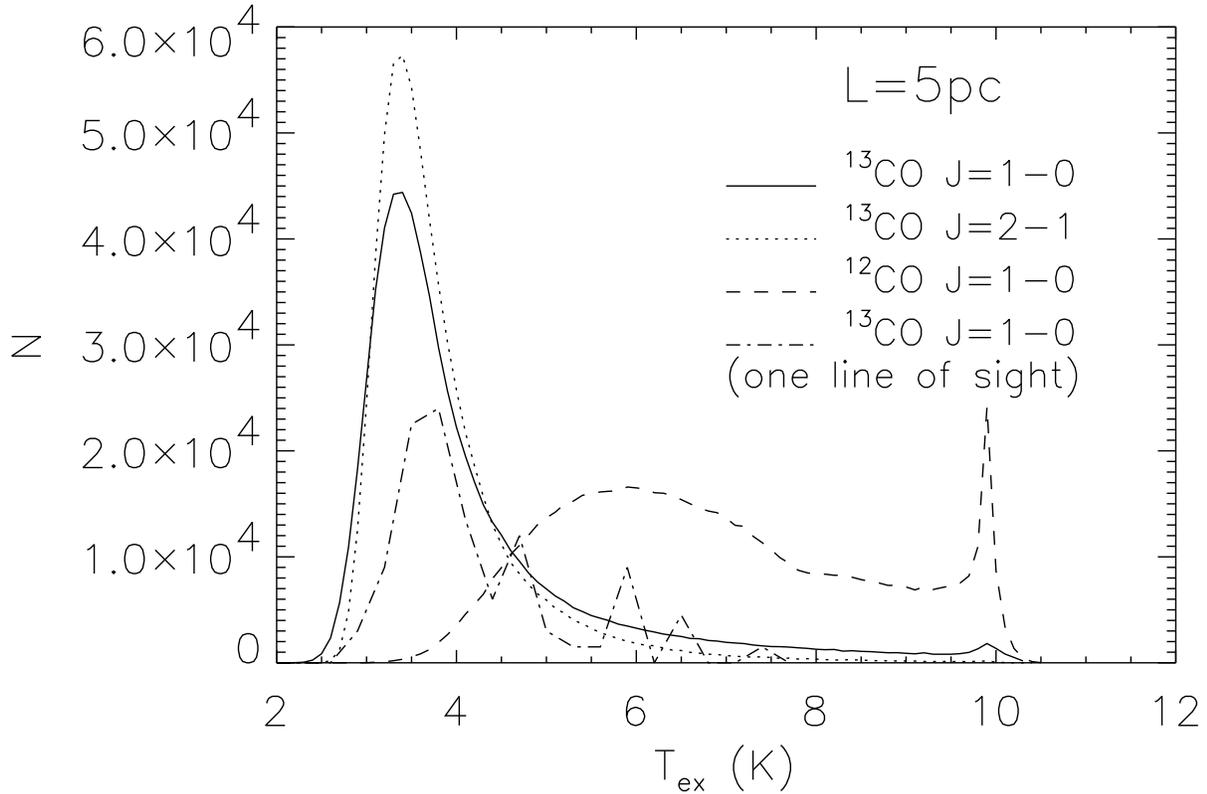}
\caption[]{Probability distribution functions for $T_{ex}$ of different transitions.
The dashed dotted line is the histogram for a single line of sight (multiplied by a
factor 1500). $T_{ex}$ of the two transitions of $^{13}$CO have rather similar distributions,
but the distribution of $T_{ex}$ of $^{12}$CO is very different, at variance with one 
of the basic assumptions of the LTE calculations. Moreover the $T_{ex}$ along a single 
line of sight is not uniform (as assumed in the LTE calculations), but has a broad
distribution.}
\end{figure}

\newpage
\begin{table}
\begin{tabular}{|c|c|c|c|c|c|}
\hline
{\bf L=20pc}; $<T_{ex}>6.8\pm2.0K$  &  N$_1$  &  N$_2$   &  N$_3$   &   N$_4$  &  N$_5$ \\ 
\hline\hline
$<N_i/N>$     & 0.40 $\pm$0.18 & 0.35 $\pm$0.13 & 0.35 $\pm$0.17 & 0.35 $\pm$0.20 & 0.34 $\pm$0.19   \\ 
\hline
$<N_i>/<N>$   & 0.50           & 0.37           & 0.45           & 0.49           & 0.46 \\
\hline\hline\hline
{\bf L=5pc}; $<T_{ex}>8.4\pm1.4K$  &  N$_1$  &  N$_2$   &  N$_3$   &   N$_4$   &   N$_5$  \\
\hline\hline
$<N_i/N>$     & 0.65 $\pm$0.16 & 0.56 $\pm$0.11 & 0.59 $\pm$0.15 & 0.60 $\pm$0.21 & 0.59 $\pm$0.19   \\
\hline
$<N_i>/<N>$   & 0.72           & 0.56           & 0.65           & 0.71           & 0.69 \\
\hline\hline
\end{tabular}
\caption{Ratios of different LTE estimations of column density (N$_i$) and real column 
density (N), for $^{13}$CO, J=1$\rightarrow$0. See the text for the definition of the N$_i$.}
\end{table}

\end{document}